# Optical contrast and refractive index of natural van der Waals heterostructure nanosheets of franckeite

*Patricia Gant[1], Foad Ghasemi[1,2], David Maeso[3], Carmen Munuera[4], Elena López-Elvira[4], Riccardo Frisenda[1], David Pérez De Lara[1], Gabino Rubio-Bollinger[3,5], Mar Garcia-Hernandez[4], Andres Castellanos-Gomez[4]*

[1] *Instituto Madrileño de Estudios Avanzados en Nanociencia (IMDEA Nanociencia), Campus de Cantoblanco, E-28049 Madrid, Spain.*

[2] *Nanoelectronic Lab, School of Electrical and Computer Engineering, University of Tehran, 14399–56191 Tehran, Iran.*

[3] *Departamento de Física de la Materia Condensada. Universidad Autónoma de Madrid, Madrid, E-28049, Spain.*

[4] *Instituto de Ciencia de Materiales de Madrid (ICMM-CSIC), Madrid, E-28049, Spain.*

[5] *Condensed Matter Physics Center (IFIMAC). Universidad Autónoma de Madrid, Madrid, E-28049, Spain.*

andres.castellanos@csic.es

ABSTRACT

We study mechanically exfoliated nanosheets of franckeite by quantitative optical microscopy. The analysis of transmission mode and epi-illumination mode optical microscopy images provides a rapid method to estimate the thickness of the exfoliated flakes at first glance. A quantitative analysis of the optical contrast spectra by means of micro-reflectance allows one to determine the refractive index of franckeite in a broad range of the visible spectrum through a fit of the acquired spectra to a Fresnel law based model.

Mechanical exfoliation is a very powerful technique to produce a large variety of high quality two-dimensional (2D) materials.[1] This sample fabrication process, however, typically yields randomly distributed flakes over the substrate surface with a large distribution of flake areas and thicknesses. Therefore fast, reliable, and non-destructive screening methods are crucial to identify ultrathin flakes and to determine their thickness. Optical microscopy based identification methods have proven to be very resourceful ways to find ultrathin flakes produced by mechanical exfoliation.[2–14] In fact, nowadays each time a new 2D material is isolated one of the most urgent things is to establish a correlation between the thicknesses of the exfoliated flakes and their optical contrast (in order to be used as a calibration guide to identify ultrathin flakes optically) and to determine the optimal substrates to identify ultrathin nanosheets by optical microscopy.

Franckeite is one of the latest novel layered materials added to the 2D materials family and up to now very little is known about this material.[15–18] One of the special characteristics that triggered the interest of the community on franckeite is the fact that it is one of the few known examples of a naturally occurring van der Waals heterostructure. Unlike most of the studied heterostructures (that are manually assembled layer-by-layer) franckeite, in its natural form, presents alternating $SnS_2$-like and PbS-like layers stacked on top of each other (see Figure 1), overcoming the major drawbacks of man-made van der Waals heterostructures: difficulty to align the crystal lattices of the different materials with atomic accuracy and presence of ambient adsorbates between the layers. Very recently Molina-Mendoza *et al.* demonstrated mechanical and liquid phase exfoliation of franckeite down to 3-4 unit cells and they fabricated field effect devices, near infrared photodetectors and PN junctions.[15] And also Velicky *et al*. isolated single unit cell nanosheets of franckeite and fabricated electrochemical devices and field effect devices.[16]

Ray *et al*. have also recently measured the photoresponse of franckeite devices in the visible and near-infrared part of the spectrum.[18] These works showed that franckeite nanosheets have an attractive narrow bandgap (<0.7 eV) and p-type doping and that they are very resilient upon atmospheric exposure. These characteristics makes franckeite an excellent alternative to black phosphorus which tends to degrade quickly upon air exposure.[19–22]

Here we study the thickness dependence of the optical contrast of mechanically exfoliated franckeite flakes. The aim of this work is to serve as a reference guide that could be used by other researcher to identify nanosheets of franckeite and to determine their thickness through quantitative analysis of their optical contrast. Our quantitative analysis of the thickness dependent optical contrast also allows us to determine the refractive index of franckeite in the visible range of the spectrum (to our knowledge this physical property was not reported in the literature yet) and therefore this work can be a starting point for further studies focused on the optical properties of franckeite nanosheets.

Franckeite flakes are prepared by mechanical exfoliation of bulk franckeite crystals extracted from a mineral rock (San José mine, Oruro (Bolivia)). The bulk franckeite crystal has been previously characterized by scanning tunnelling microscopy/spectroscopy, transmission electron microscopy, X-ray diffraction, X-ray photoemission, UV-VIS-IR absorption spectroscopy and Raman spectroscopy. More details about this characterization can be found in Ref. [15]. The flakes are firstly exfoliated onto a polydimethylsiloxane (Gelfilm by Gelpak®) carrier substrate and then transferred to a $SiO_2$/Si substrate by means of an all dry transfer technique.[23] We employed two different nominal $SiO_2$ thicknesses (~90 nm and ~290 nm) to probe the role of the $SiO_2$ thickness on the optical identification process. We selected those thickness values because they are the most standard $SiO_2$ thicknesses in the graphene and other 2D materials research. Prior to the study of the optical properties of the franckeite nanosheets, we experimentally verify the thickness of the $SiO_2$ capping layers of each employed substrate by means of reflectance spectroscopy (see the Supporting Information for more details).

Figure 2a shows a transmission mode optical image of a franckeite flake exfoliated onto the carrier Gelfilm substrate. Figure 2b shows an epi-illumination microscopy image of the same flake after being transferred onto the 292 nm $SiO_2$/Si substrate. The topography of the fabricated flakes is characterized by atomic force microscopy (AFM) to determine their thickness (see Figure 2c). Below Figure 2(a-c) we include a colour chart obtained from the analysis of tens epi-illumination microscopy images of franckeite flakes with different thicknesses. This chart can be used as a coarse guide to estimate the thickness of franckeite flakes on 292 nm $SiO_2$ substrates at first glance. Figure 2d to 2f shows similar information as Figure 2(a-c) but for flakes transferred onto a 92 nm $SiO_2$/Si substrate. Below Figure 2(d-f) we include another colour chart valid for quick identification of franckeite flakes on 92 nm $SiO_2$ substrates.

Another method to estimate the thickness of the exfoliated flakes can be obtained from the quantitative analysis of the transmission mode images, acquired on the Gelfim carrier substrate prior to the transfer. Figure 3 shows the transmittance extracted from the red, green and blue channel of the digital images where a monotonic thickness dependence of the intensity of each channel can be observed. This trend can be used as an additional way to estimate the thickness of the exfoliated flakes. On top of the plot we include a colour chart with the thickness dependent apparent colour in transmission mode images to facilitate a coarse thickness determination.

We use micro-reflectance spectroscopy to quantitatively characterize the optical contrast of franckeite flakes of different thicknesses transferred to $SiO_2$/Si substrates.[24,25] The sample is illuminated in epi-illumination mode with the white light coming from the tungsten halogen lamp of a metallurgical microscope and the light reflected from an area of the sample of 2 μm in diameter is collected and studied with a spectrometer fiber coupled to the trinocular of the microscope. We address the readers to Ref.[24] and to the Supporting Information for more details about the experimental setup and technique.

By measuring the light reflected by the bare $SiO_2$/Si substrate ($I_s$) and by the flake laying on the $SiO_2$/Si substrate ($I_f$) one can determine the optical contrast, defined as[2]:

$$C = \frac{I_f - I_s}{I_f + I_s} \ .$$

Figure 4 shows some optical contrast spectra acquired on franckeite flakes with different thicknesses transferred onto a 92 nm $SiO_2$/Si substrate. From the spectra shown in Figure 4 one can extract the thickness dependence of the optical contrast at a fixed illumination wavelength. Figure 5 shows six examples of these contrast *vs.* thickness plots, extracted for 450 nm, 500 nm, 550 nm, 600 nm, 650 nm and 700 nm illumination wavelengths. Each of these spectra can be fitted to a Fresnel law based model that accounts for the reflections and refractions of the light beam at each interface (air/franckeite, franckeite/$SiO_2$ and $SiO_2$/Si) using as fitting parameter the complex refractive index of franckeite at that specific wavelength. By doing this process for each wavelength one can determine the refractive index of franckeite nanosheets over a wide range of the visible spectrum. See the Supporting Information for more details about the Fresnel law based model.

Figure 6a shows the determined components of the refractive index ($n$ and $\kappa$) for franckeite. Note that, to our knowledge, this information was not available in the literature yet and it results crucial to further analysis of the optical properties of a material. For example, knowing the refractive index of a 2D material allows determining the substrate that optimizes its optical identification. This is done by calculating the optical contrast of a flake with a given thickness (e.g. ~1.8 nm that corresponds to a single-unit cell of franckeite) as a function of the illumination wavelength and $SiO_2$ thickness (see Figure 6b). For franckeite we found that the $SiO_2$ thickness values that optimizes the optical contrast at 550 nm of wavelength (where the human eye performance is better[26]) are 75 nm, 260 nm and 450 nm.

CONCLUSIONS

In summary, we presented a study of the optical identification of franckeite that is intended to be used as a guide for other researchers working on exfoliated franckeite. Our results allow one to determine the thickness of franckeite flakes from the analysis of their optical contrast. A deeper analysis also provides a way of determining the refractive index of franckeite in the visible spectrum, which can be a highly valuable information for further optical studies.

ACKNOWLEDGEMENTS

We acknowledge funding from the European Commission under the Graphene Flagship, contract CNECTICT-604391,from EU Horizon 2020 research and innovation program under grant agreement No. 696656 (GrapheneCore1-Graphene-based disruptive technologies), from the FP7 ITN MOLESCO (project no. 606728), from the MINECO (FIS2015-67367-C2-1-P, MAT2014-57915-R, MDM-2014-0377 and Ramón y Cajal 2014 program RYC-2014-16626), from the Comunidad de Madrid (MAD2D-CM Program (S2013/MIT-3007)) and NANOFRONTMAG-CM program (S2013/MIT-2850). RF acknowledges support from the Netherlands Organisation for Scientific Research (NWO) through the research program Rubicon with project number 680-50-1515.

FIGURES:

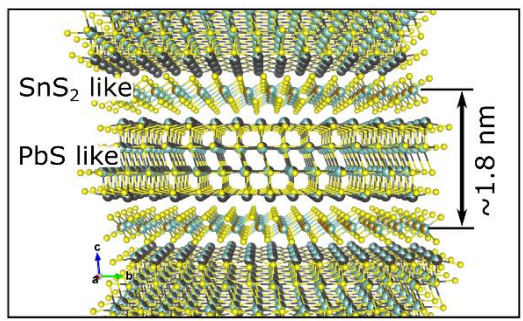

Figure 1. Sketch of the crystal structure of franckeite where the two different stack layers, the $SnS_2$-like and the PbS-like, can be seen.

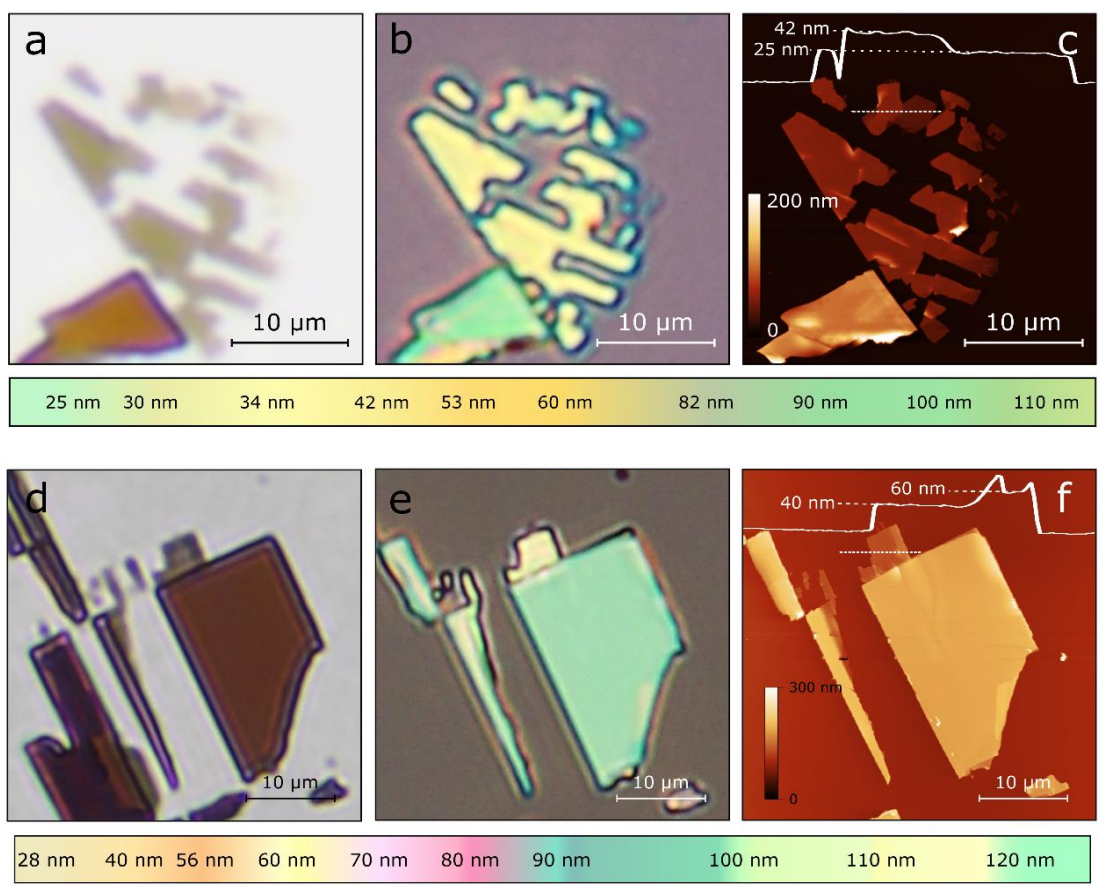

Figure 2. (a) Transmission mode optical microscopy image of franckeite flakes on a Gelfilm carrier substrate. (b) Epi-illumination optical microscopy image of the same franckeite flake after being transferred onto a 292 nm $SiO_2$/Si substrate. (c) Atomic force microscopy image of the same flake to determine its thickness. Below (a) to (c) the colour chart shows a coarse guide to determine the thickness of franckeite flakes on 292 nm $SiO_2$/Si substrates through their apparent colour. (d) to (f) Similar as (a) to (c) but for a franckeite flake transferred onto a 92 nm $SiO_2$/Si. Below (d) to (f) the colour chart shows a coarse guide to determine the thickness of franckeite flakes on 92 nm $SiO_2$/Si substrates through their apparent colour.

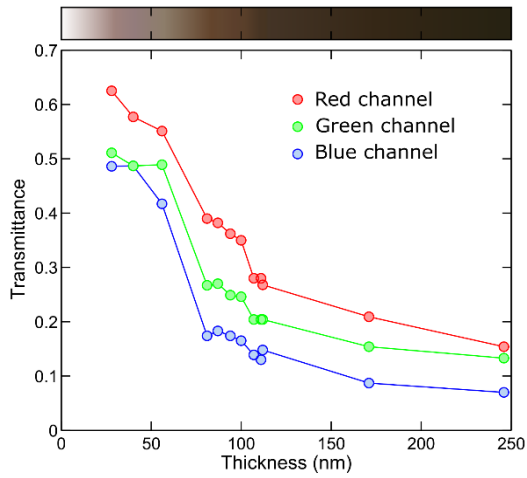

Figure 3. Thickness dependence of the transmittance acquired from transmission mode optical images of franckeite flakes on Gelfilm carrier substrates prior their transfer to SiO$_2$/Si substrates. The top colour chart shows a coarse guide to determine the thickness, from 0 nm to 250 nm, of franckeite flakes from their apparent colour in transmission mode optical images under white light.

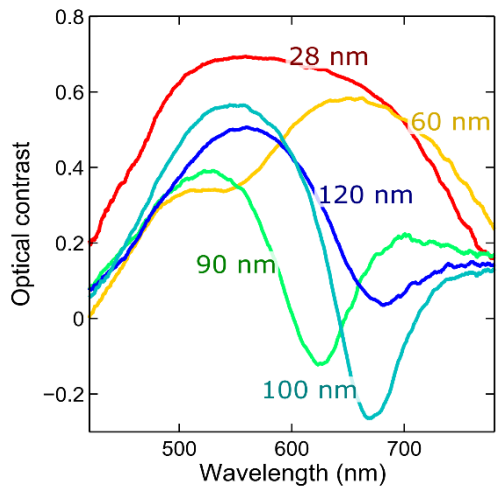

Figure4. Optical contrast spectra acquired for franckeite flakes transferred onto 92 nm SiO$_2$/Si substrates with different thickness.

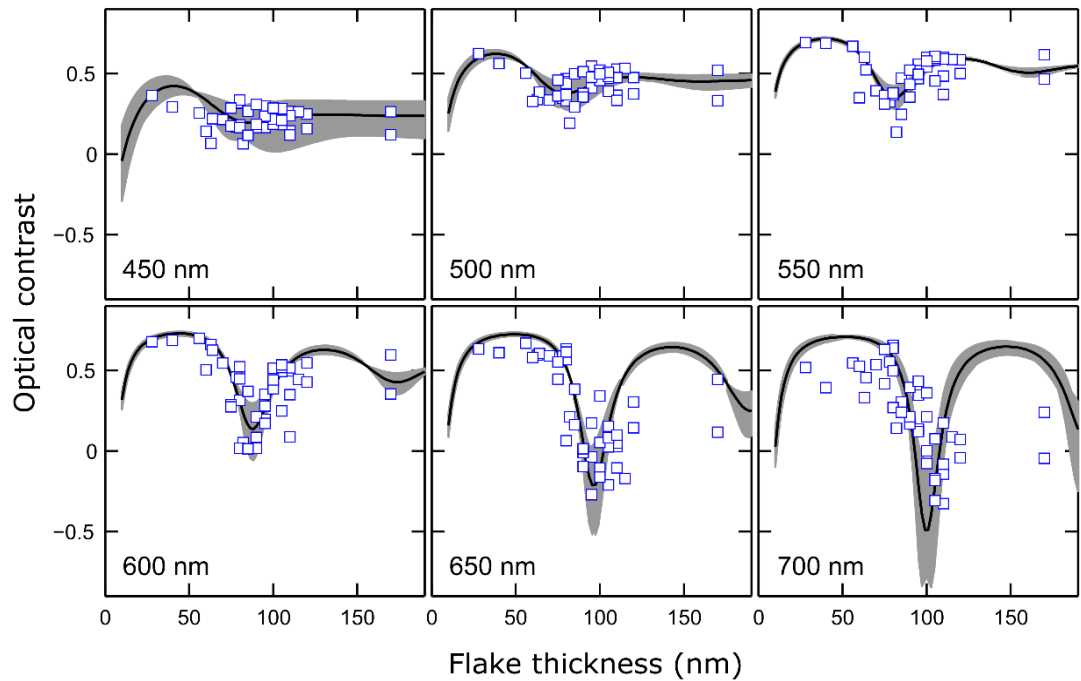

Figure5. Thickness dependent optical contrast of franckeite flakes on 92 nm SiO$_2$/Si substrate for 450 nm, 500 nm, 550 nm, 600 nm, 650 nm and 700 nm illumination wavelength. The datapoints are extracted from optical contrast spectra like those in Figure 4. The solid lines are fits to a Fresnel law based model using the franckeite refractive index as fitting parameter. The shadowed region corresponds to the uncertainty of the fit.

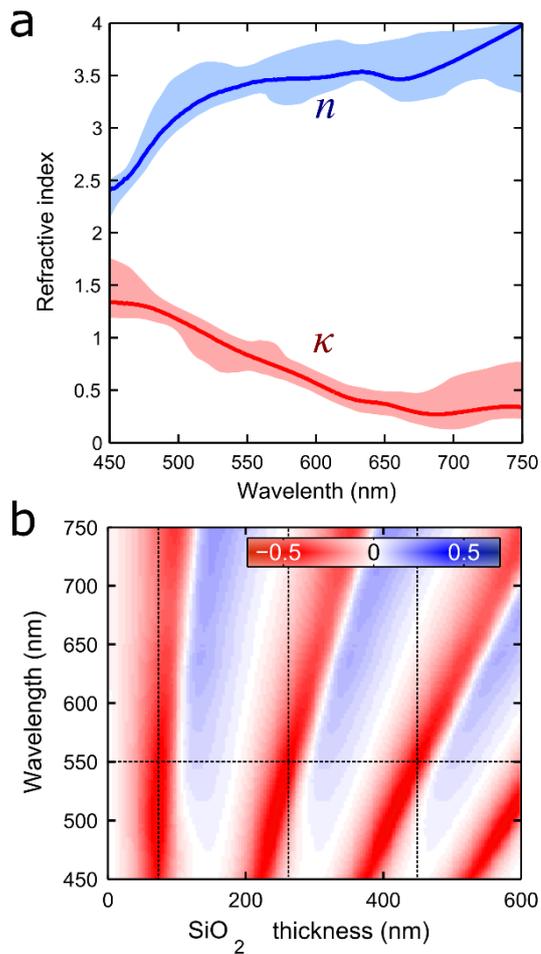

Figure6. (a) Wavelength dependent refractive index (both real and complex part) of franckeite, determined from the fit of thickness dependent optical contrast traces to a Fresnel law based mode. The shadowed region is the uncertainty of the refractive index extracted from the analysis of different datasets. (b) Calculated optical contrast for a single-layer franckeite flake as a function of the illumination wavelength and the $SiO_2$ thickness to determine the optimal $SiO_2$ capping layer to facilitate the identification of franckeite thin layers.

# Supporting Information:

# Optical contrast and refractive index of natural van der Waals heterostructure nanosheets of franckeite

*Patricia Gant[1], Foad Ghasemi[1,2], David Maeso[3], Carmen Munuera[4], Elena López-Elvira[4], Riccardo Frisenda[1], David Pérez De Lara[1], Gabino Rubio-Bollinger[3,5], Mar Garcia-Hernandez[4], Andres Castellanos-Gomez[4]*

[1] *Instituto Madrileño de Estudios Avanzados en Nanociencia (IMDEA Nanociencia), Campus de Cantoblanco, E-28049 Madrid, Spain.*

[2] *Nanoelectronic Lab, School of Electrical and Computer Engineering, University of Tehran, 14399–56191 Tehran, Iran.*

[3] *Departamento de Física de la Materia Condensada. Universidad Autónoma de Madrid, Madrid, E-28049, Spain.*

[4] *Instituto de Ciencia de Materiales de Madrid (ICMM-CSIC), Madrid, E-28049, Spain.*

[5] *Condensed Matter Physics Center (IFIMAC). Universidad Autónoma de Madrid, Madrid, E-28049, Spain.*

andres.castellanos@csic.es

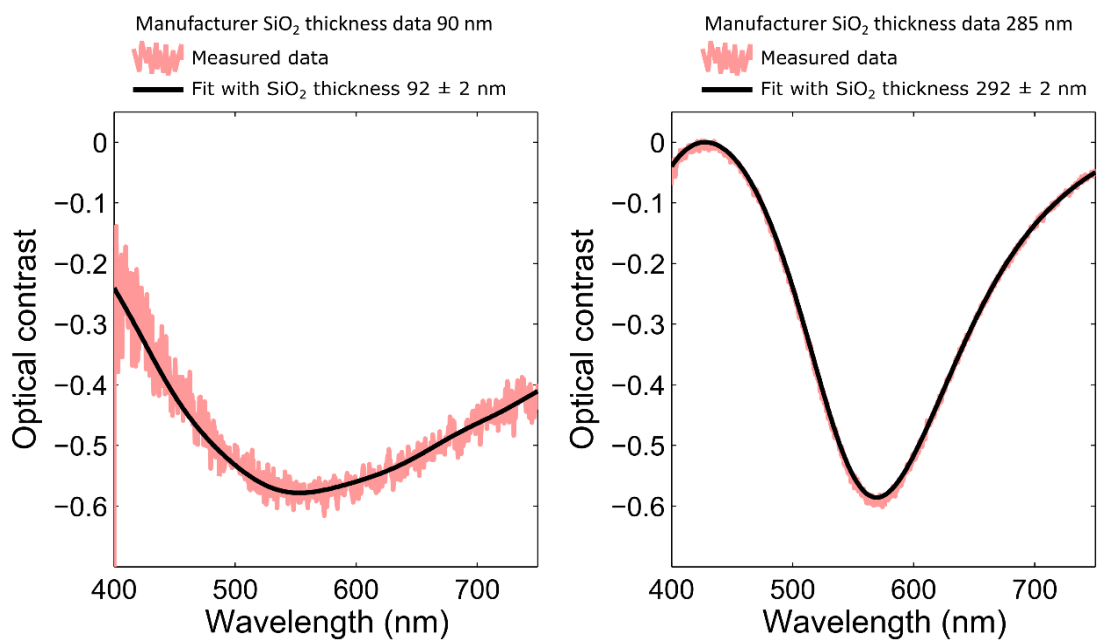

Figure S1. Determination of the SiO$_2$ thickness through the analysis of the reflectance spectrum. The optical contrast is determined by measuring the reflected light on two regions of the substrate: one where the SiO$_2$ has been previously etched (Si) and another one with the pristine SiO$_2$/Si. The two selected SiO$_2$/Si wafers have been studied (one with 90 nm of SiO$_2$ and another with 285 nm according to the manufacturer). We determined a SiO$_2$ thickness of 92 ± 2 nm and 292 ± 2nm respectively.

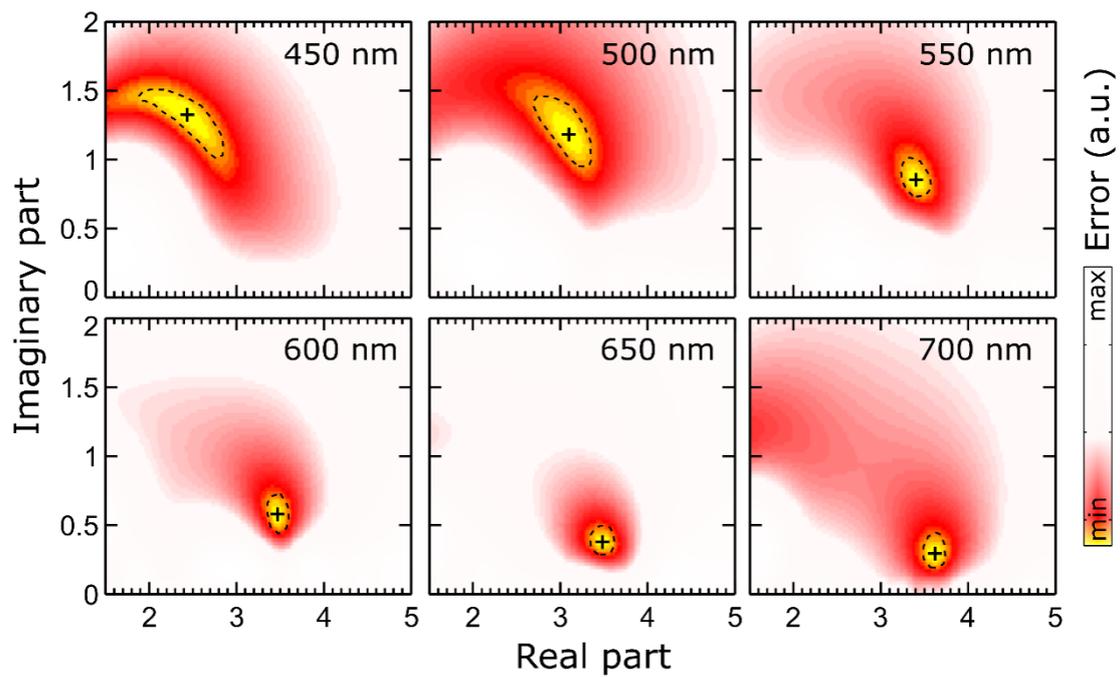

Figure S2. The colour maps show the squared error as a function of the real and imaginary part of the refractive index. The minimun squared error (cross) provides the best fit and the dashed lines show the uncertainty of the fit.

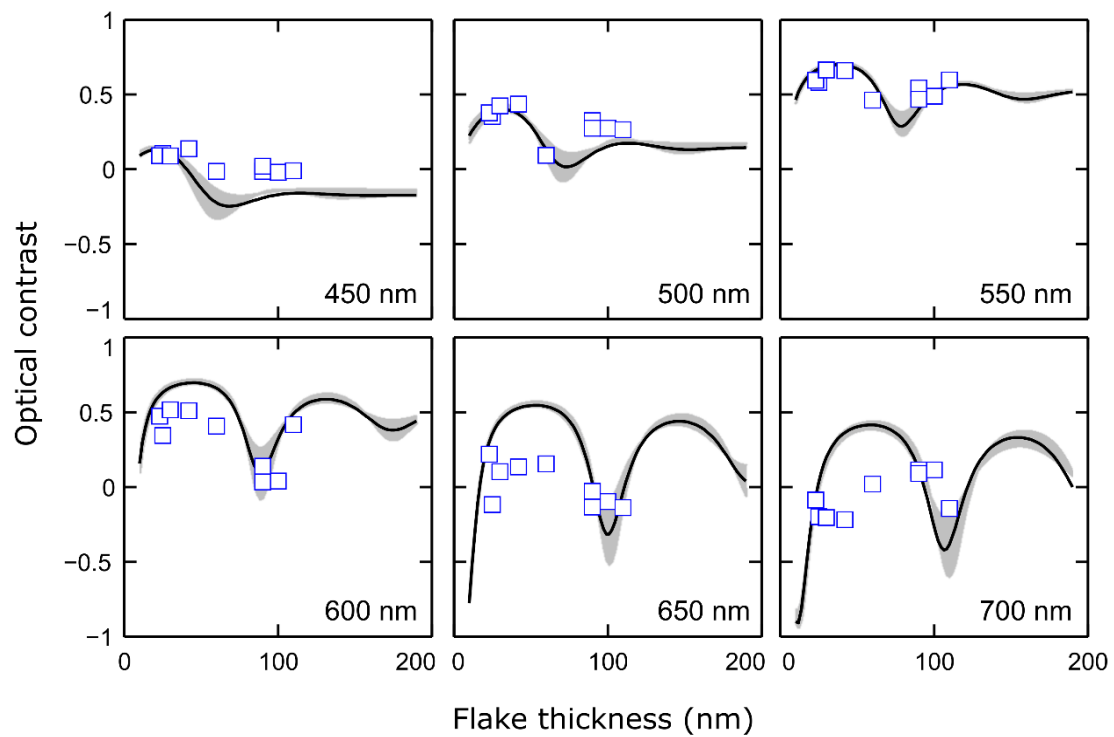

Figure S3. Same as Figure 5 of the main text but for franckeite flakes transferred to a 292 nm SiO$_2$/Si substrate.

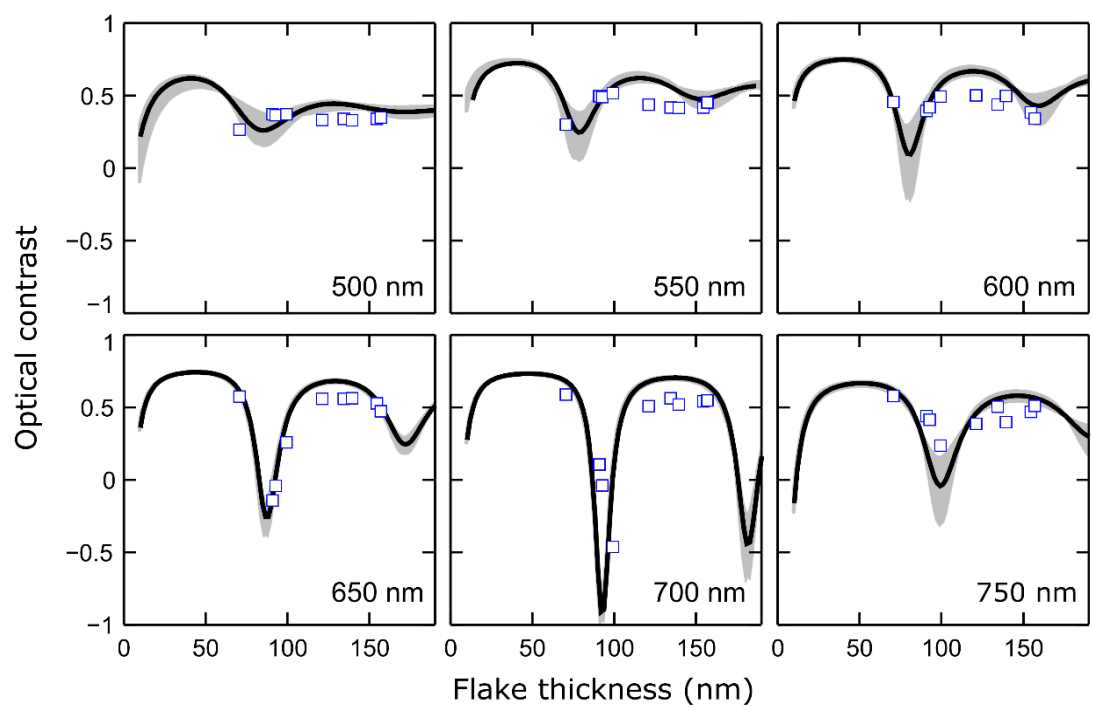

Figure S4. Same as Figure 5 of the main text for franckeite flakes transferred to a 92 nm SiO$_2$/Si substrate, but measured with a different experimental setup.

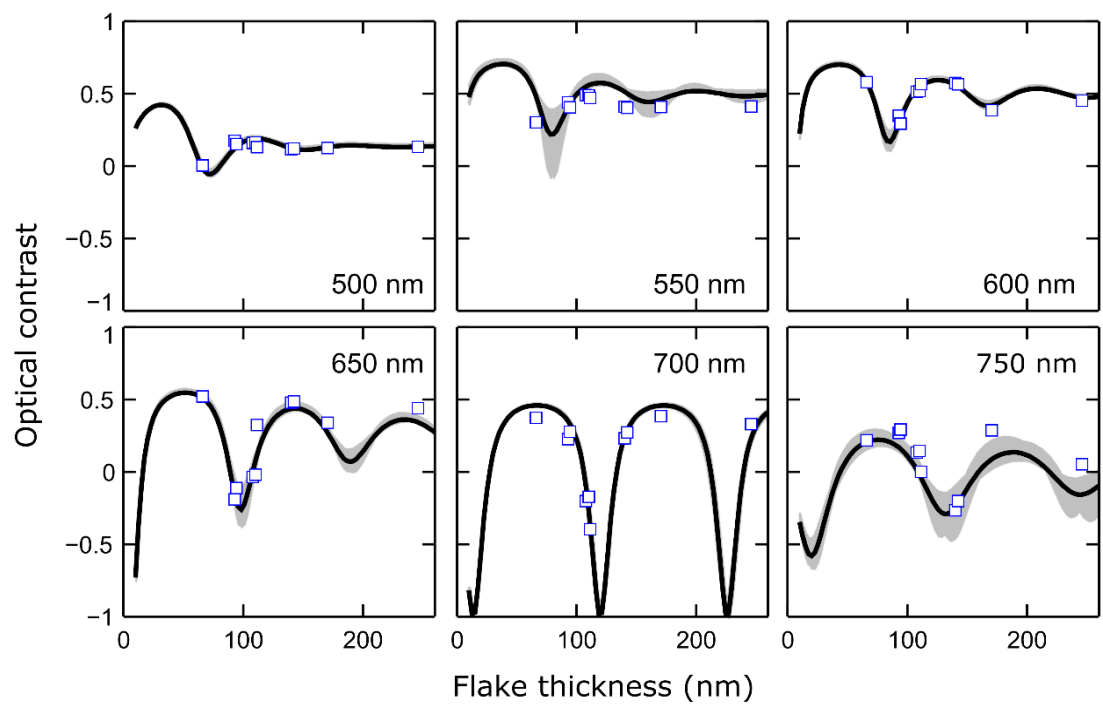

Figure S5. Same as Figure 5 of the main text for franckeite flakes transferred to a 292 nmSiO$_2$/Si substrate, but measured with a different experimental setup.

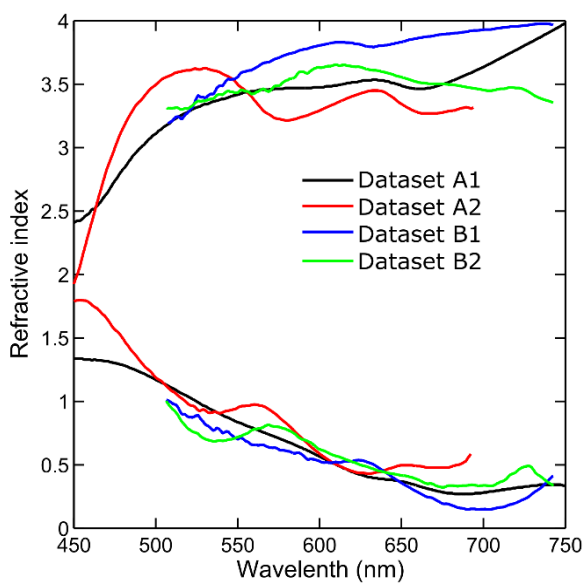

Figure S6. Refractive index determined from the fit of thickness dependent optical contrast traces to a Fresnel law based mode. Four different datasets have been analysed.

| Dataset | Microscope | SiO$_2$ thickness (nm) | Spectrometer | Number of flakes | Thickness range (nm) | Figure |
|---|---|---|---|---|---|---|
| A1 | Motic BA310 Met (50x 0.55 NA) | 92 | Thorlabs CCS200/M | 42 | 28 - 170 | Fig. 5 |
| A2 | | 292 | | 9 | 23-110 | Fig. S3 |
| B1 | Nikon Eclipse LV100 (50x 0.55 NA) | 92 | Thorlabs CCS175/M | 9 | 70-157 | Fig. S4 |
| B2 | | 292 | | 10 | 66-246 | Fig. S5 |

Table S1. Description of the different datasets analysed.